\documentclass{article}
\usepackage{bbold}
\usepackage[utf8]{inputenc}
\usepackage{natbib}
\usepackage{graphicx}
\usepackage{authblk}
\begin{document}

\title{Plant root growth against a mechanical obstacle: The early growth response of a maize root facing an axial resistance agrees with the Lockhart model} 






\author[1]{Manon Quiros}
\author[2]{Marie-Béatrice Bogeat-Triboulot}
\author[3]{Etienne Couturier}
\author[1]{Evelyne Kolb}
\affil[1]{PMMH, CNRS, ESPCI Paris, Université PSL, Sorbonne Université, Université de Paris, F-75005, Paris, France}
\affil[2]{Université de Lorraine, AgroParisTech, INRAE, UMR Silva, 54000 Nancy, France}
\affil[3]{Laboratoire Mati\`ere et Syst\`emes Complexes, Universit\'e Paris Diderot CNRS UMR 7057, 10 Rue Alice Domont et L\'eonie Ducquet, 75205 Paris Cedex 13, France}



\date{\today}

\begin{abstract}
Plant root growth is dramatically reduced in compacted soils, affecting the growth of the whole plant. Through a model experiment coupling force and kinematics measurements, we probed the force-growth relationship of a primary root contacting a stiff resisting obstacle, that mimics the strongest soil impedance variation encountered by a growing root. 
The growth of maize roots just emerging from a corseting agarose gel and contacting a force sensor (acting as an obstacle) was monitored by time-lapse imaging simultaneously to the force. \\

The evolution of the velocity field along the root was obtained from  kinematics analysis of the root texture with a PIV derived-technique. A triangular fit was introduced to retrieve the elemental elongation rate or strain rate. 
A parameter-free model based on the Lockhart law quantitatively predicts how the force at the obstacle modifies several features of the growth distribution (length of the growth zone, maximal elemental elongation rate, velocity) during the first 10 minutes. These results suggest a strong similarity of the early growth responses elicited either by a directional stress (contact) or by an isotropic perturbation (hyperosmotic bath).

\end{abstract}

\maketitle

\section{Introduction}

\subsection{Background}
Plant roots uptake the water and nutrients required to satisfy the shoot demand. They also insure the mechanical anchorage of the plant in the soil to provide a stable basis for the shoot emergence and for its resistance to external loads due for example to wind blowing, soil erosion or shallow landslides \citep{gardiner_review_2016}, \citep{stubbs_general_2019}.

The root system architecture, that is the three-dimensional spatial arrangement of the different root types, derives from branching and growth of individual roots, that continuously sense and adjust their growth according to their local environment \citep{rellan-alvarez_environmental_2016}. In addition to environmental cues like water or nutrient availability, changes in branching, growth rate or growth direction depend on the mechanical stresses experienced by the growing roots \citep{gregory_plant_2006}. In particular, the root growth velocity decays with increasing soil strength, resulting in a reduced total root length and poor above-ground development
\citep{tracy_soil_2011}. Typically for maize, the root growth velocity was reduced by 50$\%$ when the soil resistance measured with a mini-penetrometer increased from 0.2 to 2 MPa \citep{veen_influence_1990}. For all species, root elongation stops when soil strength is too large.
\\
In the current context of climate change,  the frequency of extreme wet/dry and freeze/thaw cycles can exacerbate the compaction of soils and increase their strength, thereby limiting crop yield \citep{bengough_root_2011}. As a consequence, breeding programs for plant species of agronomic interest such as wheat, soybean, rice or maize  have been developed in soil science and ecophysiology communities to identify which root traits give the better plant fitness in large strength soils \citep{lynch_future_2022}, \citep{griffiths_optimisation_2022}. In particular, some works focused on macroscopic traits such as the number of root axes or the root tortuosity in relation with soil strengths \citep{colombi_developing_2019}, \citep{popova_plant_2016}, \citep{gregory_root_2009}. However the underlying mechanisms at the root apex scale when the root tip encounters a hard pan, a compacted soil horizon or simply a rigid stone and experiences a huge resisting force, are still poorly understood. One of the main difficulty arises from getting reliable spatio-temporal information of root growth in opaque soils.\\

From a more fundamental point of view, the question arose as to quantify the maximum growth pressure, ie. the maximum axial stress a root is capable to exert on a resisting obstacle in different species. The assumption behind these studies was that roots having greater growth pressure might more easily penetrate stronger soil layers and access to water and nutrient pools. 
Interestingly the first measurements of axial force generated by growing roots were done by Pfeffer as earlier as 1893 \citep{pfeffer_druck-_1893} and reproduced long after by Gill et al. (1995)  \citep{gill_pfeffers_1955} and Souty \citep{souty_mechanical-behavior_1987}. More recently,  different techniques such as calibrated spring system, elastic beams, or digital balances have been used  (cited in Clark et al. (1999) \citep{clark_maximum_1999}) to measure root axial pushing forces in model experimental systems. By dividing the maximum force value by the root cross section, growth pressures $\sigma_{Max}$ of the order of 0.1 to 1 MPa have been obtained and were always in the range of the turgor pressure, that is the inner hydrostatic pressure inside the plant cells \citep{tracy_soil_2011}, \citep{jin_how_2013}, \citep{potocka_morphological_2018}. However the precise values of $\sigma_{Max}$ appeared to vary with the species, but also with the protocol for root growth and measurements. In particular,  $\sigma_{Max}$ depended on the way the root tip was anchored or on the location and the time of root diameter measurements \citep{kolb_physical_2017}. Indeed feedback processes due to active root responses to confining geometries could occur within different time scales.  \citep{bengough_biophysical_1997}.\\

In this context, real-time information on the growth processes are needed but until now only very few studies recorded both force and strain rate field \citep{bizet_3d_2016} and had high temporal resolution to detect potential rapid biological responses.  


\subsection{Root growth}
Primary root growth, that is root elongation, occurs in a so-called extended 'growth zone' located behind the root tip. This growth zone includes a meristematic zone where cells proliferate and an elongation zone where cells rapidly expand \citep{youssef_quantitative_2018}. The cells reaching the transition zone between the meristem and the elongation zone, leave their meristematic state and expand rapidly, increasing their volume by up to two hundred times. In addition, cell expansion is strongly anisotropic, leading to the characteristic cylindrical shape of the root. The plant cell is delimited by an hemi-permeable membrane surrounded by a rigid cell wall. The imbalance of osmotic pressures between the inside and the outside of the cell, on either side of the hemipermeable membrane results in an internal hydrostatic pressure, coined turgor pressure (or simply turgor). This pressure puts the cell wall under tension which might positively regulate cell expansion. Furthermore, the hydraulic conductivity of the cell membrane and the viscoplastic properties of the cell wall control the cell expansion rate \citep{cosgrove_wall_1987}. As modelled by Lockhart (1965) \citep{lockhart_analysis_1965}, growth is limited both by the cell wall properties and by the water uptake controlled by the hydraulic conductivity of the cell membrane \citep{cosgrove_wall_1987}.

In roots, however, the membrane hydraulic conductivity appeared not to be limiting cell expansion and thus the growth rate appears to mostly depend on the cell wall properties \citep{pritchard_control_1994}.
The cell wall is under the tension produced by the turgor pressure and deforms irreversibly when the tension exceeds a yield threshold.  Conversely the growth velocity is proportional to the turgor pressure in excess of a critical value.  


 Along the root apex, growth can be quantitatively described by the axial strain rate, also called the element elongation rate $EER$ \citep{silk_kinematics_1979} \citep{baskin_patterns_2013}. Kinematics analysis provides the velocity field and its spatial derivative gives the strain rate field \citep{silk_kinematics_1979}. Thus, kinematics experiments are a non-destructive way to obtain the local growth distribution, including the growth zone length (from the location where the local strain rate is non zero) as well as the the maximum strain rate.\\
 


\subsection{Root growth in impeding soil} 
In the presence of an impeding soil, root growth decreases with increasing soil strength. 
To include the effects of soil strength on root growth velocity, 
soil scientists proposed to consider the external resisting pressure of the soil as simply increasing the yield threshold for growth \citep{greacen_physics_1972}. If the derived phenomenological laws give the trends for growth velocity decaying with soil strength, the underlying assumptions are questionable and the force balance equations need to be rationalized. In particular it is not clear what soil stress should be incorporated inside the derived Lockhart-equations, as in situ measurements probe the soil strength resisting to the penetration of a penetrometer and do not probe the external resisting stress really experienced by the root tip during root growth. 
Moreover the parameters such as turgor pressure, cell wall extensibility or yield threshold usually involved in Lockhart models are physiological quantities that are regulated over time by the root. They were shown to depend on the mechanical stress history of the root loading \citep{bengough_biophysical_1997}. \\


In this work we address the question of how an external mechanical stress impacts the root growth rate by the use of a model experiment. We want to establish the force-velocity relationship to identify clearly how the growth process is modified with time when the root encounters a stiff obstacle and pushes against it. To answer this question, we built a new experimental setup combining force and kinematics measurements with a spatio-temporal analysis of the root growing against a force sensor acting as an obstacle. We chose a model species, maize, whose radicule has a typical diameter in the millimeter range and whose growth in the absence of mechanical stresses has been well characterized.
Our analysis not only provides the root growth velocity but also fine kinematic parameters such as the growth zone extent and the local strain rate. These kinematic parameters are essential for characterizing the growth process and confronting them to growth models such as the Lockhart law \citep{lockhart_analysis_1965}.



  %

\section{Material and Methods}
\subsection{Experimental setup}
\begin{figure}[htbp]
\begin{center}
(a)\includegraphics[scale=0.22]{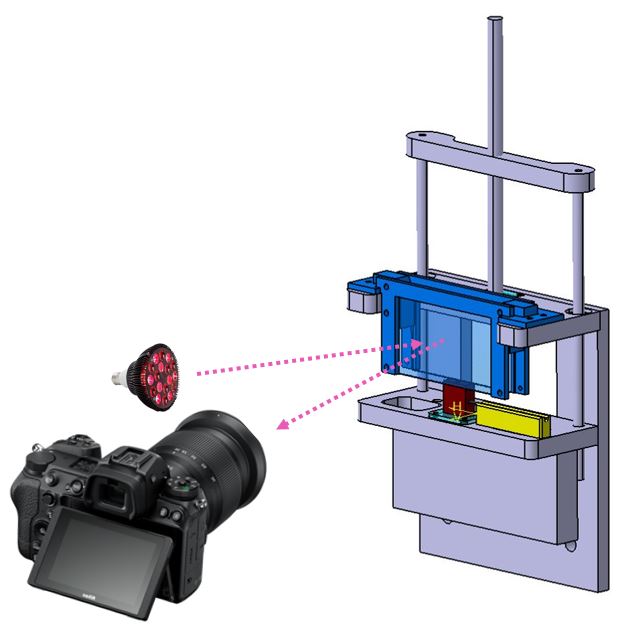}
(b)\includegraphics[scale=0.22]{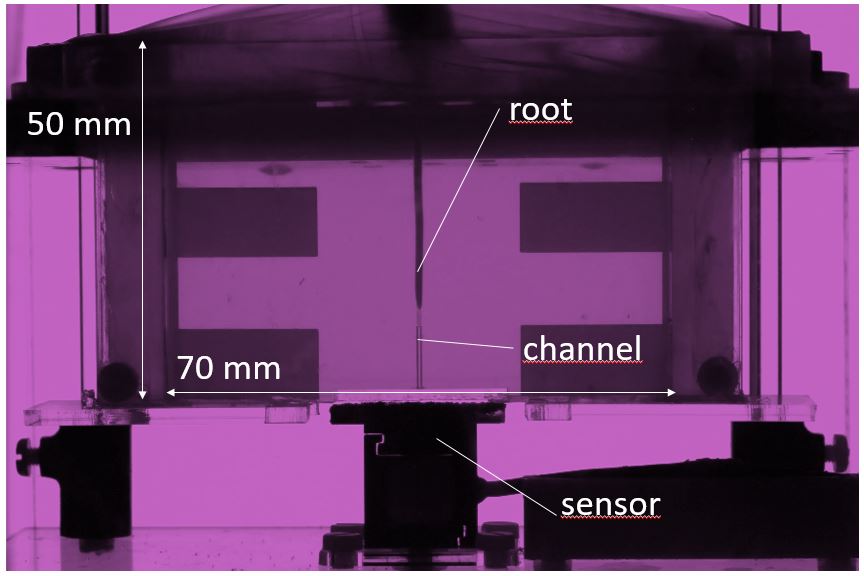}
\caption{Experimental setup. (a) Schematic view: The agarose-filled cell (blue) is fixed on a support (grey).  The vertical position of the force sensor (red) can be adjusted just at the outlet of the vertical channel guiding the root growth. A 3D-printed support (yellow) holds all the electrical wires. The input wires are connected externally (not shown) to a stabilized power supply and the output wires are connected to a LabView interface. A CCD camera saves the images of the root illuminated by an infra-red lighting.(b) Global view of the growth chamber with the transparent agarose gel filling the chamber and the maize root inside the vertical channel perpendicular to the horizontal surface of the force sensor.}
\label{fig:setup}
\end{center} 
\end{figure}

The setup is made of a growth chamber mounted on a support with a movable part allowing the vertical adjustment of the force sensor just underneath the channel the root grew in. The growth chamber was made of two parallel glass plates (height 50 mm, width 70 mm) placed vertically and held together at a distance of 10 mm by an assembly of laser-cut Plexiglass walls (Figure~\ref{fig:setup}a).\\
Agarose powder (2\% w/v, SeaKem LE Agarose) was dissolved in de-ionized water at a temperature of 90°C, until becoming transparent. A small amount of the solution was then poured in the chamber, let to cool in order to seal the bottom edges, then the rest was poured in until the chamber was full. A graphite rod of calibrated diameter (0.7 mm) slightly smaller than the root’s diameter ($\approx$0.9~mm) was inserted and maintained vertically in the middle of the cell until the agarose gel solidified. 
Afterwards, the graphite rod was removed, forming a vertical channel for the root to grow in (Figure~\ref{fig:setup}b).\\
Maize seeds (variety P9874, Pioneer) were soaked in de-ionized water, aerated thanks to a bubbler for 24 hours, then were transferred to humid paper for 24 hours and held upright, root tip facing down in the direction of gravity. Seedlings with a 0.5 to 1 cm-long radicle were selected for the experiments. The radicle was carefully inserted in the channel at the top of the chamber, and the seed was maintained in place with a parafilm sheet. Vaseline was spread on the bottom-exposed areas of the agarose gel to avoid water losses. Then the agarose-filled chamber was fixed on the support just over the force sensor.\\
The setup allowed the simultaneous monitoring of the root growth (through the agarose gel) and of the root force when the radicle emerged from the gel and pushed again the force sensor. The root grew in the channel for around 19 hours and arrived vertically onto the force sensor, acting as an obstacle. Temperature was continuously monitored and experiments were conducted with temperatures in the range of  $\theta_0 \in \lbrack 21^{\circ}C - 28^{\circ}C \rbrack$. For each experiment with a given $\theta_0$, the standard deviation on $\theta_0$ did not exceed $0.5^{\circ}C$ during the time course of kinematics measurements. In this range, the 2$\%$ w/v agarose gel behaves as a stiff elastic material with an elastic modulus of around $22.5\pm 0.5~kPa$ (measured with a HAAKE Rheometer, Thermofisher).



\subsection{Force measurements}
The force was measured thanks to a Futek (LSB200) force sensor, with a maximum range of 100 g and a stiffness of 4828 ± 5 N/m. The signal was acquired at a frequency of 1000 Hz, then averaged every second during the whole experiment duration.
In order to avoid the root tip slipping when contacting the force sensor, we fixed a rough sand-covered plexiglass rectangle on the top face of the force sensor. The sand particles (diameter $\leq$ 350 $\mu m$) were painted black to avoid any unwanted light reflections.

\subsection{Kinematics of root growth}
\subsubsection{Experimental observation and image processing}
Root growth was monitored by time-lapse photography. The whole setup was illuminated with low angled Infra-Red lighting (IR, $\lambda=850 nm$) (Figure~\ref{fig:setup}a). This IR lighting has two advantages. First, root growth is not disturbed by phototropic effects that might occur with visible light. Second, IR lighting gives a texture to the root surface, that is, a pattern of bright points that can be followed along time to get the displacement field along the root \citep{youssef_quantitative_2018}. For this kinematics study of the root growth, we used a high-resolution CCD camera (Nikon D5200, $4000 \times 6000$ pixels) whose IR absorbing filter was removed, with a macro lens (Nikkor 60 mm). The observation field was 25 mm in height with a typical resolution of $4.7\mu m$ per pixel.  When the root tip was entering the observation field  and approaching the contact, the images were taken every minute during 15 hours. 


Then, images were processed with Kymorod, a Matlab app developed by Bastien et al. (2016) \citep{bastien_kymorod_2016}, that performed PIV (Particle Image Velocimetry) on the texture of elongating organs. Kymorod was run with a time lapse of 4 minutes between images. We recovered the raw displacement fields as a function of the curvilinear abscissa $s$ along the root skeleton (Figure~\ref{fig:EER}) and used a fitting procedure to determine the local velocity $v_l$ and the strain rate (noted $EER$ for Elementary Elongation Rate) profiles.

\subsubsection{Velocity, Strain rate profiles and fitting procedures}
Numerical derivation can be done in numerous ways by deriving local fits (splines) or global fits. To minimize the number of parameters and make the $EER$ estimation the more robust possible, we chose the simplest global fit with the following constraints:  The fit function for the strain rate profile $EER(s)$ must have a finite support (for modelling the limited extent of the growth zone and the absence of growth elsewhere) and once spatially integrated, it should give a sigmoid shape similar to velocity profile. These conditions required that the fit function for $EER(s)$ had to be an isoscelese triangle :

\begin{equation}
   s\to 2a(s-b)\mathbb{1}_{[b,b+c]}+(2a(b+2c-s))\mathbb{1}_{[b+c,b+2c]}.
   \label{Triangle}
\end{equation}

where $\mathbb{1}_{[x,y]}$ is the function that is 1 in between $x$ and $y$ and 0 elsewhere and with $a, b, c$ the three fit parameters describing the triangular shape. The function family (\ref{Triangle}) stands for triangle shape displacement profile whose height is the maximum strain rate $EER_{max}=2ac$  and whose base is the length of the growth zone $L_{GZ}=2c$ (see the triangular sketch of Figure~\ref{fig_schema}).

Hence the velocity fit function family is obtained by spatial integration over $s$ of the function (\ref{Triangle}):
\begin{equation}
   s\to a(s-b)^2\mathbb{1}_{[b,b+c]}+(2ac^2-a(s-b-2c)^2)\mathbb{1}_{[b+c,b+2c]}+\mathbb{1}_{[b+2c,\infty[}2ac^2+d
\label{IntegTriangle}
\end{equation}
with 4 parameters only, one of which ($d$) being due to integration. Other velocity fit functions were proposed in the literature \citep{peters_tailor-made_2006}; this one was retained for having a finite support of the growth zone and only 4 parameters.

In this way, growth velocity $v$ of the root  is given by the maximum of the fit function (\ref{IntegTriangle}): $v=2ac^2$.




  %

\section{Experimental Results}

\subsection{Growth before contact}

The roots grew  along the vertical channel inside the agarose gel. The water required for root growth was supplied by the gel surrounding the root and a drop of water probably released by the compressed gel was observed just at the root cap, preventing any dehydration of it. The root diameter being slightly larger than the channel, there was no air interface between the root and the gel, allowing a good IR specs visualization and PIV processing. After a transient regime of 2 hours during which the root growth velocity increased, a stationary regime was reached with a typical velocity of around 4 cm/day. 


\begin{figure}[htbp]
\begin{center}
\includegraphics[scale=0.43]{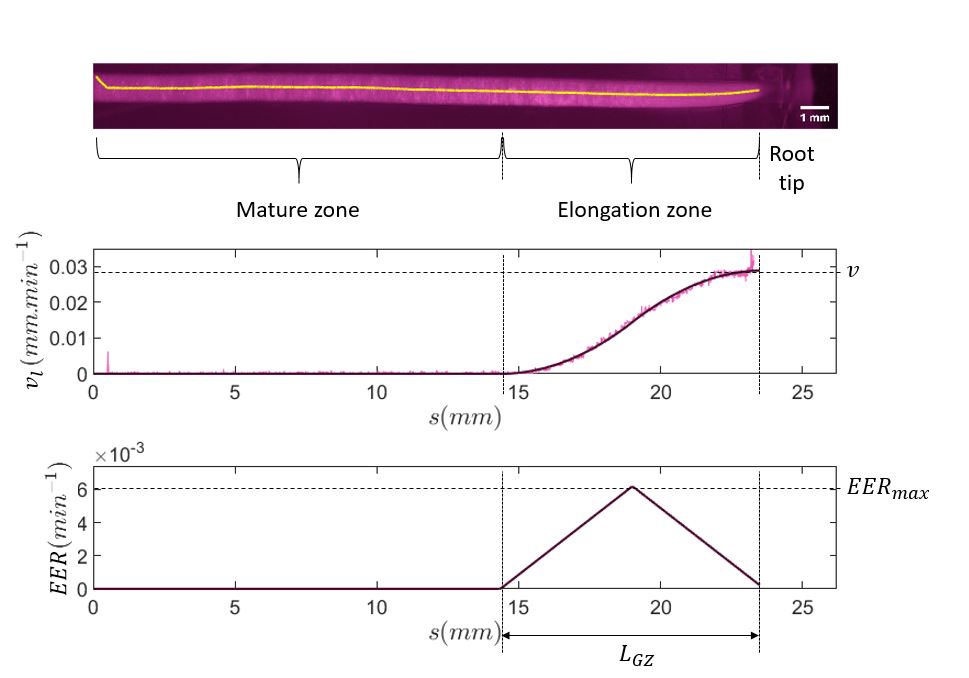}
\caption{Top panel: Cropped and 90 degrees rotated image of the root apex (before contact) under IR lighting with the yellow skeleton obtained from Kymorod analysis superimposed on it. Middle panel: Local velocity profile $v_l$ along the root curvilinear abscissa $s$ with raw data in purple and its fit in black line with its maximum, ie. the growth velocity $v$. Bottom panel: Strain rate ($EER$) along the root abscissa $s$ obtained by spatial derivation of the fitted velocity profile, with indications of $L_{GZ}$ for the length of the growth zone and $EER_{max}$ for the maximum strain rate.}
\label{fig:EER}
\end{center} 
\end{figure}

The imaging of the root with infra-red lighting and the subsequent analysis with the PIV analysis of Kymorod allowed to obtain the root skeleton (yellow line in Figure~\ref{fig:EER}) and the local velocity profile $v_l$ as a function of the curvilinear abscissa $s$ along the root length. A typical example of this profile before the root contacts the obstacle is given in Figure~\ref{fig:EER}, middle panel. The reference $s=0$ was arbitrarily set at the extreme point of the observation frame toward the seed. The velocity $v_l$ increased from the curvilinear abscissa $s=14~\textrm{mm}$ until the root tip at around $s=24~\textrm{mm}$ (purple curve of the middle panel of Figure~\ref{fig:EER}). The extent of this zone where $v_l$ departs from zero corresponds the growth zone. The raw velocity field was fitted by equation (\ref{IntegTriangle}). The fit function (black curve) nicely reproduced the experimental data. The maximum of this fit gave the root growth velocity $v = 28.9~\mu m.min^{-1}$. 
From this fitting procedure, we got the three fit parameters of equation (\ref{Triangle}) necessary to properly define the strain rate profile (bottom panel of Figure~\ref{fig:EER}) and the growth parameters. Hence for this example before contact, $L_{GZ}=8.64~\textrm{mm}$ and $EER_{max} = 6.18 \times 10^{-3}~min^{-1}$.




 We proceeded in the same way for all the roots investigated. The average growth velocity before contact was
 $v = 28.5 \pm 2.5~\mu m.min^{-1}$. The averaged growth zone length ($L_{GZ}$) was $8.9 \pm 0.6 mm$ and the averaged maximum strain rate ($EER_{max}$) was $5.9 \pm 0.6 \times 10^{-3}~min^{-1}$. Root growth velocity, growth zone length and maximum strain rates were similar to those already described in previous maize studies \citep{sharp_growth_1988}, \citep{walter_expansion_2003}, indicating that our set-up with the root inserted in gel did not induce any oxygen deficiency nor affect root growth.


\subsection{Contact}
After a typical duration of 19 hours inside the gel, the root tip reached the gel bottom and touched the force sensor with a normal incidence. The precise determination of the contact time was challenging since the zone of root-sensor contact was blurred, due to the gel bottom-air interface, preventing a visual determination of the contact occurrence.
Considering that the contact time did not necessarily coincide with the force increase, we used a method based on the spatio-temporal analysis of the displacements of the bright points of the root texture. 

\begin{figure}[htbp]
\centering
\includegraphics[scale=0.3]{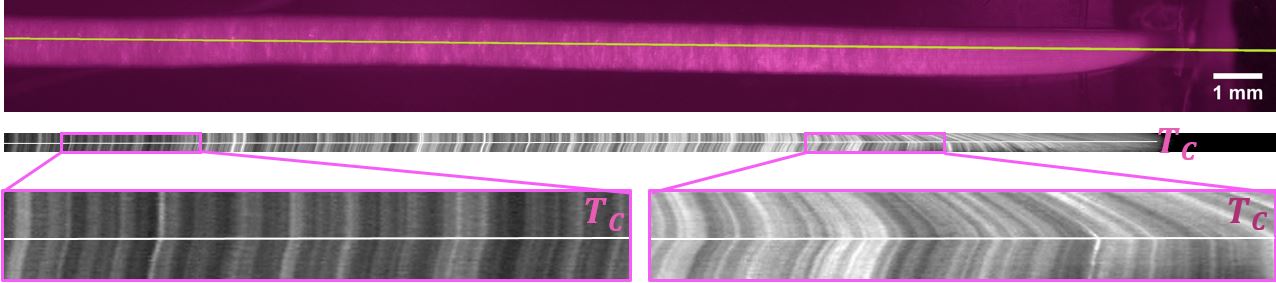}
\caption{Spatiotemporal analysis of the root images. Top: picture of the maize root under IR lighting with the 1 pixel-thick straight line used as a slice for the spatiotemporal analysis. Middle: Full spatiotemporal image converted in gray-levels with increased contrast. The horizontal axis shows variations along the root, with the root cap on the right side. The vertical axis represents increasing times from top to bottom. Bottom images: zooms of the spatiotemporal in the mature zone (left) and in the growth zone (right). The white line indicates the slope break determining the time $T_C$ when the root contacts the force sensor.}
\label{fig:spatio}
\end{figure}

We draw a vertical line of 1 pixel width along the root on each IR image and analyzed the resulting spatio-temporal pattern. In Figure~\ref{fig:spatio} the image of the root has been rotated by 90 degrees. In the middle and bottom panels each horizontal line corresponds to one slice along the root at a given time, and the vertical from top to bottom corresponds to increasing times, each slice being separated by 1 min. On the right part of the slices, the growth zone can be identified by the location of oblique lines (see the right zoom of the spatiotemporal in Figure~\ref{fig:spatio}): the progressive shift of the bright spots shows the displacement of cells along the root axis. The slope of these oblique lines increases from left to right due to the cumulative effect of local growth on the more apical cells of the root tip. In the mature zone (left zoom of Figure~\ref{fig:spatio} ) where there is no growth, the bright points of the texture stay in place along time before contact, resulting in vertical lines in the spatio-temporal image. At time $T_C$, a slope break is observed in the vertical lines, the spots appear displaced backward toward the seed. As the force sensor was much more rigid than the root, the contact led to a backward movement of the IR specks that could result from a shortening of the mature zone due to compression or to a pullback of the whole seedling. We thus considered this event as due to the contact of the root tip with the force sensor. To determine $T_C$, isointensity lines of the spatio-temporal diagram were detected using the Contour function of Matlab; the longest ones (going from the initial to the final times of the spatio-temporal diagram) were selected; a piece-wiese linear function with two pieces was fitted; the time $T_C$ corresponded to the junction between these two pieces. Note that a second noticeable slope break occurred later when the root markedly bends.




\subsection{Force build-up}
\begin{figure}[htbp]
\centering
\includegraphics[scale=0.45]{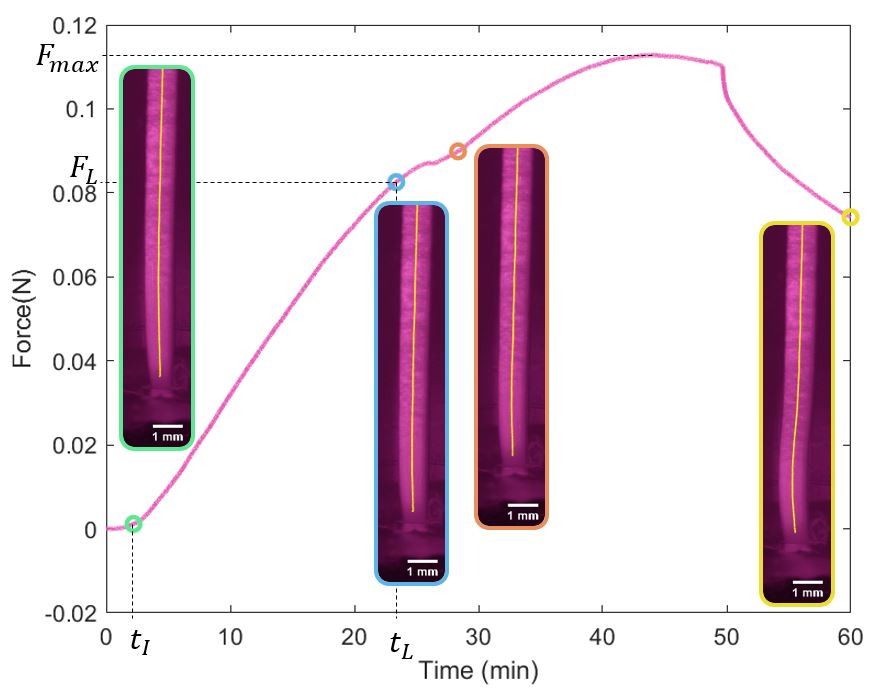}
\caption{Force $F$ as a function of time $t$. $t=0$ corresponds to the contact time. Characteristic times are indicated with dotted vertical lines: time $t_I$ (resp. $t_L$) is the beginning (resp. end) of the linear increase of force. Insets: pictures of the root at different characteristic times ($t$= 2.5, 23.4, 28.3 and 60 min from left to right respectively).}
\label{fig:force_time}
\end{figure}

 Once the contact time was determined, we could precisely follow the evolution of force $F$ as a function of the rescaled time $t= T-T_C$, for which we set $F(t=0)=0$. A typical evolution of the force $F$ exerted by the root on the force sensor is plotted as a function of the rescaled time $t$ (Figure~\ref{fig:force_time}) with insets corresponding to the images of the root and the computed skeletons (in yellow) at different characteristic times. \\
 The force was observed to increase with time, as the root continued to grow and thus to push against the force sensor. From time $t_I$ which denotes the time where the force started to increase noticeably, the signal evolved in a linear way until time $t_L$. The end of this linear regime was obtained mathematically following the successive steps below: \begin{itemize}
 \item Calculating linear fits for the portions of the $F$ versus $t$ curve beginning at $t_I$ and ending at the successive times $t$.  
\item Calculating the successive quadratic distances between the portions of curve $F$ versus $t$ and the corresponding fits. Each quadratic distance is normalized by the corresponding $t$.
\item  Plotting the histogram of quadratic distances and identifying its first peak. This first peak corresponds to all the data where the linear fit was very close to the experimental force-time data. The location and width of this first peak are calculated by the matlab function findpeaks (MinPeakHeight=40 and MinPeakProminence=4). $t_L$ is given by the maximal time whose quadratic distance lays in the first peak of the histogram.
\end{itemize}

 By this method, we determined for the typical example of Figure~\ref{fig:force_time}, a time $t_L=23.4~\mathrm{min}$ for a time $t_I=2.5~\mathrm{min}$ and a typical slope $\frac{\Delta F}{\Delta t}=3.9~\mathrm{mN/min} $. When normalized by the initial root growth velocity $V_0=28.9~\mathrm{\mu m/min}$ we obtained a value of $\frac{\Delta F}{\Delta t \times V_0}=k_{eff}=135~\mathrm{N/m}$, which has the dimension of an effective root stiffness ($k_{eff}$). 


 Then after time $t_L$, the signal of force versus time rounded off until a maximum force value of around $F_{max}=0.11$~N was reached for this root. After this maximum force $F_{max}$ was reached, there was a marked and spatially-extended bending of the root: the root axis appeared curved along a typical length of $7.5 \pm 0.7$ mm (right inset of Figure~\ref{fig:force_time}). This event was thus associated to a macroscopic buckling of the root inside the gel.\\
 The existence of a linear regime of force increase and then a rounding off of the force signal until a maximum force value were observed for all investigated roots. When averaged over $n=7$ roots, the duration of the linear regime was $t_L-t_I=15 \pm 4~\mathrm{min}$ and the characteristic slope was $4.5 \pm 0.7 ~\mathrm{N/min}$ corresponding to an effective stiffness $k_{eff}=159 \pm 32~\mathrm{N/m}$ (the value following $\pm$ is the standard deviation over $n=7$ roots).

\subsection{Growth response}
\begin{figure}[htbp]
\centering
\includegraphics[scale=0.5]{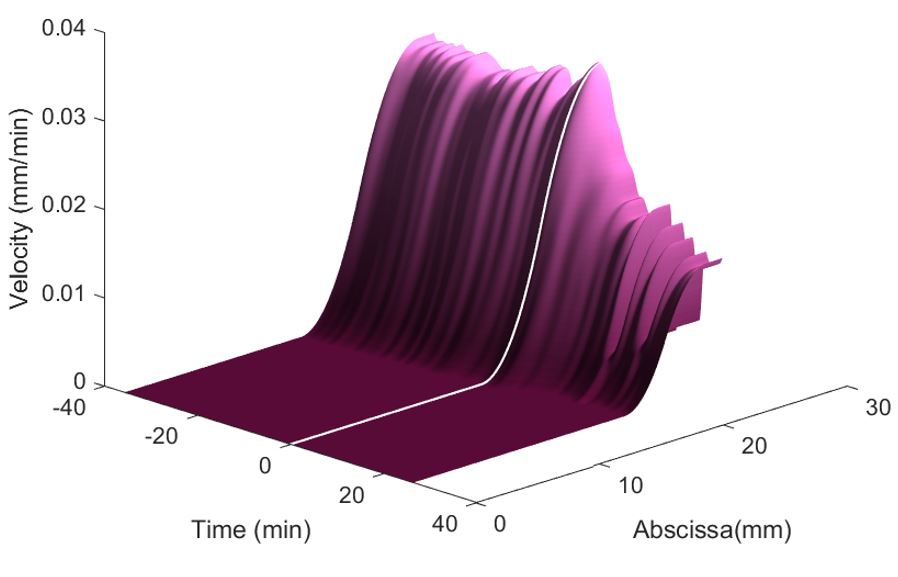}
\caption{3D map of the velocity profile $v_l$ along the curvilinear abscissa $s$ of the root skeleton for different times $t$ before and after the contact (identified by the white line at $t=0$).}
\label{fig:nappe}
\end{figure}

From the successive images and Kymorod analysis, we could follow the kinematics of root growth before and during the contact with the force sensor. The velocity profiles varying with time are represented with a 3D map in Figure~\ref{fig:nappe} for the same root as in Figure~\ref{fig:force_time}. The white line is the velocity profile at the rescaled time $t=0$ corresponding to the contact with the force sensor ($T_C$). Before contact, that is for times $t < 0$, the successive velocity profiles were similar with a growth zone starting at around $s=14~\mathrm{mm}$ and extending until $s=24~\mathrm{mm}$. Small variations of the maximum velocity (that is the root growth velocity $v$) are visible during the 30 min preceding the contact with a value of $v$ in between 0.0278 mm/min and 0.031 mm/min. Once the root tip contacted the force sensor, we observed a drastic change of behaviour. The maximum velocity decayed rapidly over 15 minutes and then more gradually over the next 15 minutes. 
Growth monitoring by kinematics allowed to highlight that the root continued to grow although its tip was blocked by the rigid force sensor. This could not have been shown by a more classical method such root tip displacement monitoring. This also implies that the increase in root length was "dispersed" by either seed pullback, mature tissue compression or micro-bendings. 
 
\subsection{Coupling Force and Growth}
\begin{figure}[htbp]
\centering
\includegraphics[scale=0.5]{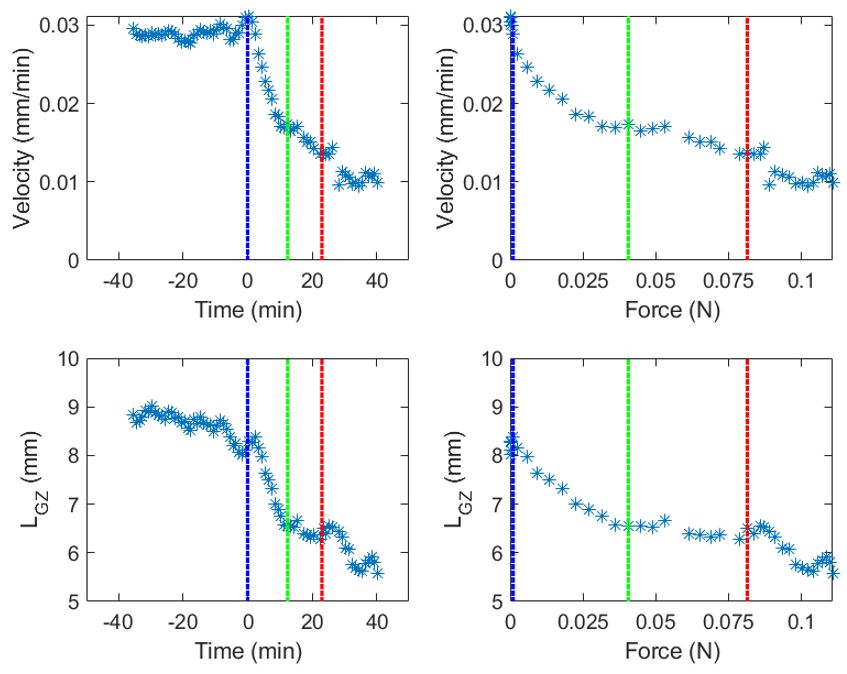}
\caption{Growth parameters, ie. growth velocity $v$ (top panels) and growth zone length $L_{GZ}$ (bottom panels) obtained by fitting the velocity profiles $v_l(s)$ with equation (\ref{IntegTriangle}) for the same root as in Figure~\ref{fig:nappe}. Left panels: as a function of time $t$, $t=0$ corresponding to the contact time (dotted line). Right panels: as a function of measured force $F$. The blue, green and red dotted lines represent respectively the contact time, the time corresponding to $F=F_{N}=0.04 N$, and the time $t_L$ when the linear rise in force ends.}
\label{fig:LGZcontact}
\end{figure}

The fitting procedures were applied to all the velocity profiles before and after the contact and gave the growth  parameters shown in Figure~\ref{fig:LGZcontact}. In particular, the growth velocity $v$ was plotted as a function of time (top left panel in Figure~\ref{fig:LGZcontact}). In the typical example, we could observe a quasi-linear decay of the growth velocity $v$ over a duration of around 10 minutes after contact. Then the decay was slower and the growth did not stop in the represented time range. Note that we stopped the kinematic fitting procedure when the root clearly bent at time $t_B=40~\mathrm{min}$. This bending did not necessarily occur in the observation plane, which does not allow to use the kinematics analysis beyond this time. The simultaneous acquisitions of force and IR images also allowed to plot the growth velocity as a function of force (top right panel in Figure~\ref{fig:LGZcontact}). After a marked decay of velocity with increasing force until an amplitude of 0.04~N, the growth velocity seemed to decay much slower. Growth persisted even if the resisting force was still increasing.

Besides growth velocity, the fitting procedures also gave the growth zone length $L_{GZ}$ (lower panels of Figure~\ref{fig:LGZcontact}). Starting from $L_{GZ}= 9~\mathrm{mm}$ well before contact, the growth zone length shrank rapidly to 6.5~mm after around 10~min then decayed more slowly. In a similar manner as the growth velocity, $L_{GZ}$ also decayed with force, seemed to stabilize at 6.5~mm for a force level of 0.03-0.04~N, before a small rise and a further decrease down to 5.5 ~mm.

\begin{figure}[htbp]
\centering
\includegraphics[scale=0.4]{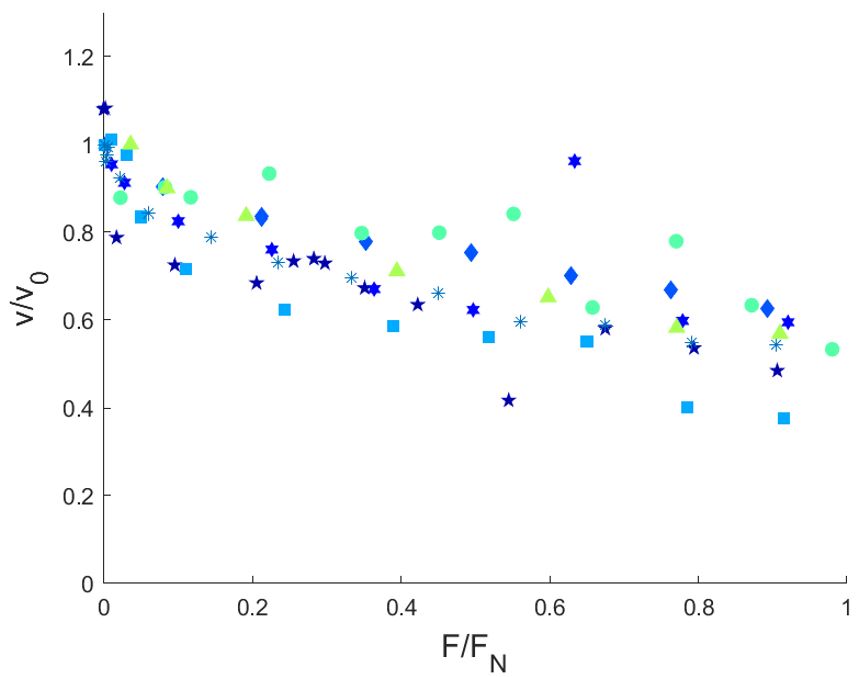}
\caption{Velocity normalized by the initial velocity $V_0$ just before contact as a function of the rescaled force $F/F_N$ where $F_N=0.04~N$ for the 7 different roots (one given symbol and color per root, the asterisks corresponding to the root described in the preceding figures).}
\label{fig:v_F_Renormalise}
\end{figure}

We proceeded to the same analysis for force and growth for the different roots and summarised in Figure~\ref{fig:v_F_Renormalise}. The growth velocity was normalized by its value $V_0$ just before contact and the force $F$ was divided by a constant force level of $F_N=0.04~N$ corresponding to the levelling off of the velocity in the typical example. Despite the inherent biological variability of the seeds, the curves of the rescaled velocity $V/V_0$ versus rescaled force $F/F_N$ collapse rather well for the $n=7$ roots we measured.


\section{Lockhart's law dictates root-obstacle interaction at short time scale}
\subsection{Model}
We propose to interpret our experimental results in the framework of Lockhart law \citep{lockhart_analysis_1965}. This law establishes a relationship between the strain rate at the cell scale and the turgor pressure inside the cell that puts the rigid cell wall under tension. The cell wall growth regulation by turgor follows the same numerical law as a Bingham fluid which deforms irreversibly above a yield stress : while turgor pressure $P$ exceeds a given threshold in pressure, cell walls flow and expand. We recall that whereas $P$ is isotropic, the primary cell walls are mechanically anisotropic. 



\begin{figure}[h!]
\centering
\includegraphics[scale=0.4]{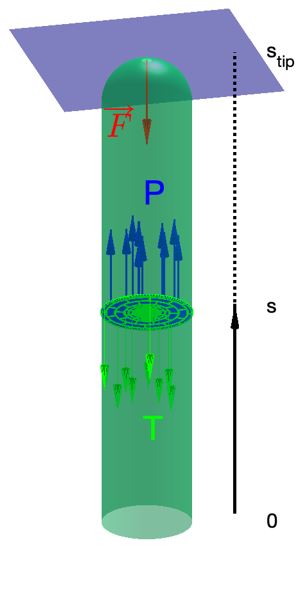}
\includegraphics[scale=0.3]{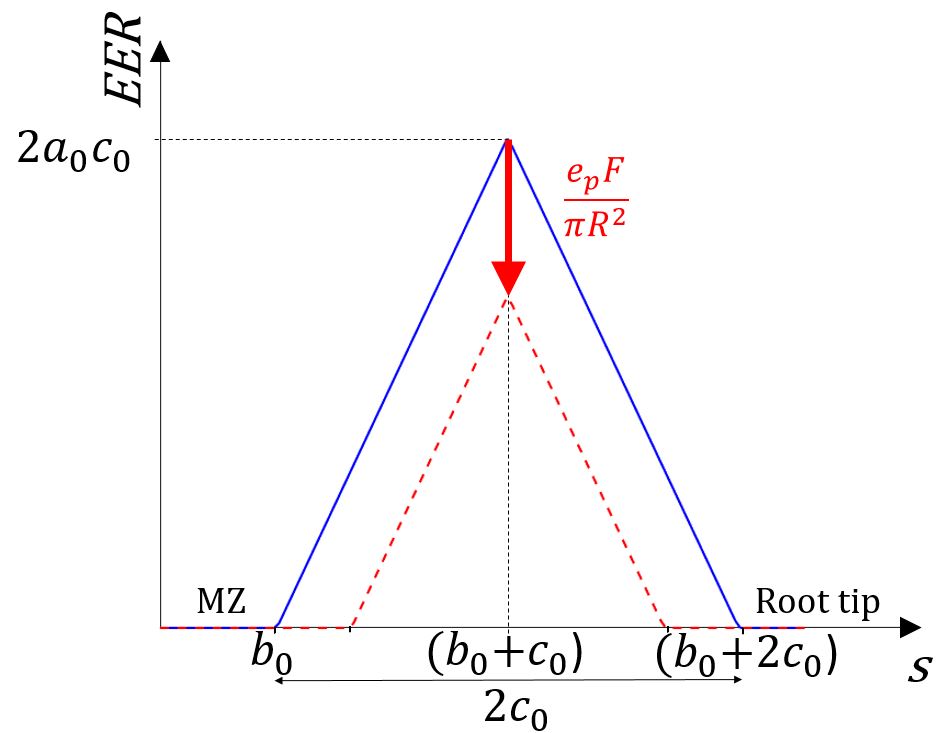}
\caption{Sketch of a root. $s$ is the curvilinear abscissa along the root. In the transversal cross-section at $s$ : $\mathcal{A}_{cytoplasm}$ (resp. $\mathcal{A}_{wall}$ ) stands for the cross-sectional area of the cytoplasm (blue zone) (resp. cell wall (green zone)). $P$ stands for the turgor pressure, $T$ stands for the cell wall tension and $F$ stands for the apical force.}
\label{fig_schema}
\end{figure}

We have adapted this approach to the case of a root encountering an obstacle for modelling the growth-force relationship. Notations and analysis are inspired by the work of Dyson et al. (2014) \citep{dyson_mechanical_2014}. $F$, the force exerted at the root tip by the obstacle is balanced by the contributions of $P$, the turgor pressure, and $T$, the cell wall tension. Other environmental forces such as the frictional forces on the root flanks are neglected. The force balance applied to the root part laying between the $s$ cross-section and the tip (See Figure~\ref{fig_schema}) gives:

\begin{equation}
-F-\int\displaylimits_{wall} T \cdot \mathrm{d} A_w +\int\displaylimits_{cytoplasm} P \cdot \mathrm{d} A_c =0.
\label{Force_balance}
\end{equation}

The area of the $s$ cross-section $\mathcal{A}(s)$ being $\pi\left[ R(s)^2 \right ]$ can be decomposed in its cytoplasmic part  $\mathcal{A}_{cytoplasm}(s)$ (liquid part under turgor pressure $P$, in blue in the the left panel of Figure~\ref{fig_schema} and its cell-wall part  $\mathcal{A}_{wall}(s)$ (solid part under tension $T$ in green in Figure~\ref{fig_schema} left):

$$\mathcal{A}(s)=\mathcal{A}_{wall}(s)+\mathcal{A}_{cytoplasm}(s).$$

Using $\bar{T}$, the tension averaged on the cell wall of the $s$ cross-section and $\bar{P}$, the turgor averaged on the cytoplasmic area of the $s$ cross-section, equation~(\ref{Force_balance}) rewrites:

$$-\mathcal{A}_{wall} \bar{T}+\mathcal{A}_{cytoplasm} \bar{P}-F=0.$$

$\bar{T}$ is a simple function of $\bar{P}$ and $F$:
\begin{equation}
\bar{T}=\frac{\mathcal{A}_{cytoplasm}\bar{P}-F}{\mathcal{A}_{wall}}.
\label{eq_tension}
\end{equation}

Neglecting the tension variations over the $s$ cross-section, the Lockhart equation expressing the strain rate $EER(s,F)$ for an applied force $F$ reads:

\begin{equation}
\begin{centering}
EER(s,F)=e_T(s) \Big(\bar{T}-Y_T(s)\Big)_+
\label{eq_Lockhart_base}
\end{centering}
\end{equation}

 $e_T$ (resp. $Y_T$) being the local extensibility (resp. the local threshold) expressed in tension. The subscript '+' indicates that the formula is valid if $\Big(\bar{T}-Y_T(s)\Big) >0$ and $EER(s,F)=0$  otherwise. Substituting (\ref{eq_tension}) in (\ref{eq_Lockhart_base}) gives:

\begin{equation}
EER(s,F)=e_T(s) \Big(\frac{\mathcal{A}_{cytoplasm}}{\mathcal{A}_{wall}}\bar{P}-Y_T(s)-\frac{F}{\mathcal{A}_{wall}}\Big)_+.
\label{eq_Lockhart_Area}
\end{equation}

The strain-rate profile before contact $EER(s,0)$ is given by the same equation~(\ref{eq_Lockhart_Area}) by setting $F=0$. Then it is possible to rewrite the strain rate during contact $EER(s,F)$ in the following way:

\begin{equation}
EER(s,F)=\Big(EER(s,0)-\frac{e_T(s)F}{\mathcal{A}_{wall}}\Big)_+.
\end{equation}

The extensibility $e_T$ has been estimated indirectly by retrieving data points from Frensch and Hsiao (1995) who were studying the maize root growth submitted to water stress. Their data in Figure 7 of \citep{frensch_rapid_1995} showed the extensibility to decrease slowly from tip to base (Figure 7D) : the extensibility expressed in turgor was $e_P=2.36\pm 0.7 \mathrm{MPa^{-1} h^{-1}} \ (N=3)$, relative standard deviation $0.0545\ (N=3)$, whereas the threshold profile followed a bell shape inversely correlated with the $EER$ profile (relative standard deviation $0.14\ (N=3)$). These observations led us to suppose $e_P$ to be constant along the growth zone and to explain the strain rate variation solely by the threshold variation:

\begin{equation}
EER(s,F)=\Big(EER(s,0)-\frac{e_PF}{\mathcal{A}_{cytoplasm}}\Big)_+,
\end{equation}

$e_T$ being converted in $e_P$ according to:

$$e_T=(\mathcal{A}_{wall}/\mathcal{A}_{cytoplasm})e_P.$$

Thus the strain rate profile at a given force $F$ can be simply expressed as a linear combination of the force $F$ and the strain-rate profile before contact:
 
\begin{equation}
EER(s,F)=\Big(EER(s,0)-\frac{e_PF}{\pi R^2}\Big)_+,
\label{eq_final}
\end{equation}

where $\mathcal{A}_{wall}(s)$ is neglected compared to $\mathcal{A}_{cytoplasm}(s)$, leading to $\mathcal{A}_{cytoplasm}(s)\approx\pi\left[ R(s)^2 \right ]$.\\

Substituting the fit of $s\to EER(s,0)$ with formula~(\ref{Triangle}) in equation~(\ref{eq_final}) gives:

\begin{equation}
     EER(s,F)=\Big(2a_0(s-b_0)\mathbb{1}_{[b_0,b_0+c_0]}+(2a_0(b_0+2c_0-s))\mathbb{1}_{[b_0+c_0,b_0+2c_0]}-e_P F/(\pi R^2)\Big)_+.
\end{equation}

The function $s\to EER(s,F)$ calculated with this formula is still triangle shaped with a height (Figure~\ref{fig_schema} right):
\begin{equation}
 EER_{max}=2a_0c_0-e_P F/(\pi R^2).
\end{equation}

\begin{figure}[h!]
\centering
\includegraphics[scale=0.45]{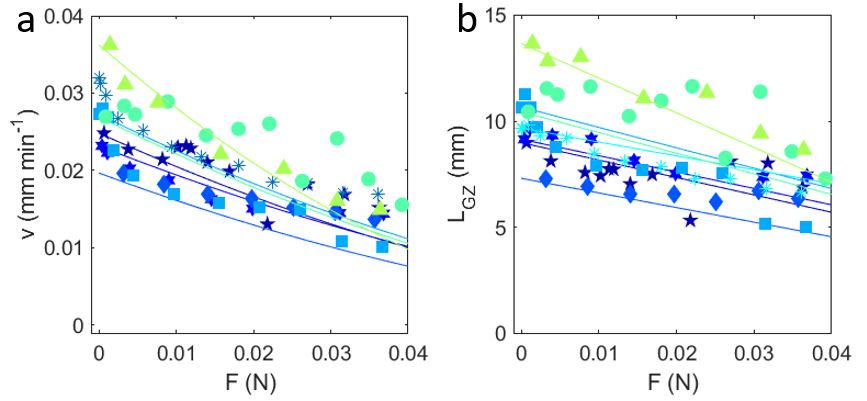}
\caption{a. Force vs velocity. b. Force vs growth zone length. For both a and b, the circles are experimental points and the continuous lines are the theoretical predictions.}
\label{fig_theorie}
\end{figure}

Then the growth zone length ($L_{GZ}$) corresponding to the basis of the triangle $s\to EER(s,F)$ is easily obtained by noting that $s\to EER(s,F)$ and $s\to EER(s,0)$ are two similar triangles:
\begin{equation}
L_{GZ}=2c_0\frac{2a_0c_0-e_PF/(\pi R^2)}{2a_0c_0}
\end{equation}
which simplifies in:
\begin{equation}
L_{GZ}=2c_0-\frac{e_PF}{a_0\pi R^2}.
\end{equation}
The growth velocity is given by the area of the triangle ($s\to EER(s,F)$)
\begin{equation}
v=\frac{(2a_0c_0-e_P F/(\pi R^2))^2}{2a_0}.
\end{equation}
After substitutions with the parameters before contact $L_{GZ,0}$, $EER_{max,0}$ and $v_0$, the kinematic parameters after contact are: 
\begin{equation}
L_{GZ}=L_{GZ,0}\left(1-\frac{e_PF}{\pi R^2EER_{max,0}}\right).\label{formule_GZ}
\end{equation}
and
\begin{equation}
v=v_0\left(1-\frac{e_PF}{\pi R^2EER_{max,0}}\right)^2.\label{formule_vitesse}
\end{equation}
Both the velocity and the growth-zone length can be predicted (Formula (\ref{formule_GZ}) and (\ref{formule_vitesse})) from the displacement profile before impact, the root radius and the force. The match between prediction and experimental data is good (Figure \ref{fig_theorie} a and b). 
Moreover the linear regression coefficient of the curve $F$ vs
$(\pi R^2EER_{max,0})\left(1-\sqrt{v/v_0}\right)$ (resp. $F$ vs
$
(\pi R^2EER_{max,0})\left(1-L_{GZ}/L_{GZ,0}\right)
$ lead to estimations of $e_P$ both very close to each other ($1.87\pm 0.36~\mathrm{MPa^{-1} h^{-1}}$ and $1.87\pm0.52~\mathrm{MPa^{-1} h^{-1}}$, $N=7$) and close from the measurements of \citep{frensch_rapid_1995} $2.36~\mathrm{MPa^{-1}h^{-1}}$.


\subsection{Discussion}
Mechanical cues trigger lots of responses at different time scales \citep{landrein_connected_2019}; the present data show that the first phase (ten first minutes or $F<0.04N$ between the blue and green vertical dotted lines on Figure \ref{fig:force_time}) of the root growth response to an obstacle can be described with evolution laws derived from the Lockhart model. In this model, the strain rate ($EER$) remained proportional to the wall tension (above a predetermined threshold), while the wall tension was decreasing due to the increasing force exerted at the root tip. After this first phase, the root growth velocity decreased more slowly than expected from the model while the force kept increasing and finally reached a plateau (see Figure \ref{fig:LGZcontact} between the green and red vertical dotted lines). 
 In a third phase (after the red line of Figure~\ref{fig:LGZcontact} or after $t_L$ in Figure~\ref{fig:force_time}), the force was no more linear with time and the velocity-force relationship was noisy. We interpret this phase as due to a localized bending allowing the dispersion of new tissue (produced by growth) transversely to the vertical axis. Note that the macroscopic bending corresponding to a clear buckling event happened later on and kinematics analysis were stopped there.
 

In the past, most tests of the Lockhart growth law were non-directional (i) by varying cell internal pressure either using a cell pressure probe \citep{zhu_enlargement_1992} or  varying external osmolarity \citep{frensch_rapid_1995}, or (ii) by applying an external pressure by use of a pressure chamber \citep{cosgrove_wall_1987} while monitoring growth. The reduction of cell wall tension by hyper-osmotic treatments was shown to elicit two homeostatic responses of the plant: a reduction of the growth threshold to come back to the initial growth rate coined "cell wall loosening" \citep{green_metabolic_1971} and an increase in internal osmolyte concentration to at least partially restore turgor, coined "osmotic adjustment". For instance, a maize root with an initial turgor of 0.67 MPa immersed in a 0.3 MPa mannitol solution (hyper-osmotic bath) ceased to grow very rapidly when the turgor decreased below 0.6 MPa \citep{frensch_transient_1994} suggesting an initial very high turgor threshold. Turgor reached a minimum (0.34 MPa) two minutes later after which it started to recover for 30 min by osmotic adjustment. Finally the growth was restored ten minutes later when the turgor was only 0.46 MPa, indicating cell wall loosening and a new decreased turgor threshold (0.46~MPa in comparison with 0.6~MPa). In a second experiment, after a lower osmotic shock (0.1 MPa), the turgor threshold was decreased in two minutes. In addition, the growth never ceased but after an initial drop the initial growth rate and turgor were re-established in less than 10 min \citep{frensch_rapid_1995} indicating that osmotic adjustment (that restores turgor) took place in a short time scales, again within a frame of 10 minutes.

In our experiment, the recorded wall tension drop induced by contact forces inferior to $F_{N}=0.04\ N$ corresponds to a turgor drop lower than $0.05\ MPa$ (calculated with $F_{N}/\pi R^2)$. According to \citep{frensch_rapid_1995}, we can suppose that an osmotic adjustment 
 could have occurred after 10 minutes, increasing turgor and at least partially cancelling the effect of the contact force on the root growth velocity. Osmotic adjustment in response to impedance/external force was already evidenced in Greacen and Oh (1972) \citep{greacen_physics_1972} and Atwell and al. (1988) \citep{atwell_physiological-responses_1988}). Cell turgor pressure was shown to be strongly increased in response to growth blockage by an axial force \citep{clark_complete_1996} but without indication on the short term dynamics. Turgor pressure was however not measured and  untangling the role of cell wall loosening versus osmoregulation would call for experiments where force is maintained constant by controlling the displacement of the obstacle during the growth. The transition toward longer term mecanoperceptive responses which also tend to slow-down growth \citep{coutand_biomechanical_2000} could also be studied.

Directional methods to test Lockhart growth by stretching organs with small weights were first developed as an alternative to measure plastic deformation associated to growth of soybean stem \citep{nonami_wall_1990} and proved to be numerically equivalent to non-directional methods \citep{nonami_primary_1990}. The method was refined to estimate easily both yield threshold and extensibility of maize leaves after exposure to salinity \citep{cramer_kinetics_1991}. Compression experiments of stem pieces (coined "External Force method") were carried out by Cosgrove \citep{cosgrove_wall_1987} to study the dynamics of the turgor-growth relationship but "the pattern of force were highly variable" and the technique non pursued. In the light of our experiments it was probably due to the variability of the buckling threshold. Our study is thus the first proof of the equivalence between non-directional and directional methods for plants in the case of compression: parameters estimated with a non-directional method \citep{frensch_rapid_1995}, an hyperosmotic treatment, can predict quantitatively the response of the $EER$ distribution ($L_{GZ}$ and $EER_{max}$) to a directional solicitation, the contact with an obstacle (Figure \ref{fig_theorie}). This model illustrates the power and limitations of the analogy between cell wall growth and rheology to make predictive models of plant tissues with complex growth patterns: the yield threshold distribution inferred from the pre-contact kinematics and the extensibility are sufficient parameters to describe the ten first minutes of the interactions with an obstacle.

\section{Conclusion}
As a conclusion, we built a model experimental system to study how an external mechanical stress impacts the primary growth of the maize radicule. We coupled force and kinematics measurements to investigate the first stages of the root apex contacting a stiff obstacle. We established the force-velocity relationship and characterized fine kinematics parameters such as the growth zone extent and the maximum strain rate. We proposed a derived Lockhart model to take into account the compression force produced by the axial growth against the obstacle.  Through this model and by using parameters of the literature for maize roots submitted to water stresses (non-directional methods), we could predict without any adjustable parameter the decrease of velocity and growth zone lengths with force, within the first ten minutes of contact. These results suggest a strong similarity of the early growth responses elicited either by a directional stress (contact) or by an isotropic perturbation (hyperosmotic bath).

\section{Funding}
M.B. Bogeat-Triboulot was supported by a grant overseen by the French National Research Agency (ANR) as part of the "Investissements d'Avenir" program (ANR-11-LABX-0002-01, Lab of Excellence ARBRE). E. Couturier was supported by AnAdSpi ANR-20-CE30-0005-01. 


\bibliographystyle{plainnat}

\bibliography{References_20220328_2.bib}

\begin{thebibliography}{44}
\providecommand{\natexlab}[1]{#1}
\providecommand{\url}[1]{\texttt{#1}}
\expandafter\ifx\csname urlstyle\endcsname\relax
  \providecommand{\doi}[1]{doi: #1}\else
  \providecommand{\doi}{doi: \begingroup \urlstyle{rm}\Url}\fi

\bibitem[Atwell(1988)]{atwell_physiological-responses_1988}
Bj~Atwell.
\newblock Physiological-{Responses} of {Lupin} {Roots} to {Soil} {Compaction}.
\newblock \emph{Plant and Soil}, 111\penalty0 (2):\penalty0 277--281, October
  1988.
\newblock ISSN 0032-079X.
\newblock \doi{10.1007/BF02139953}.
\newblock (doi : 10.1007/BF02139953).

\bibitem[Baskin(2013)]{baskin_patterns_2013}
Tobias~I. Baskin.
\newblock Patterns of root growth acclimation: constant processes, changing
  boundaries.
\newblock \emph{Wiley Interdisciplinary Reviews-Developmental Biology},
  2\penalty0 (1):\penalty0 65--73, February 2013.
\newblock ISSN 1759-7684.
\newblock \doi{10.1002/wdev.94}.
\newblock (doi : 10.1002/wdev.94).

\bibitem[Bastien et~al.(2016)Bastien, Legland, Martin, Fregosi, Peaucelle,
  Douady, Moulia, and Hoefte]{bastien_kymorod_2016}
Renaud Bastien, David Legland, Marjolaine Martin, Lucien Fregosi, Alexis
  Peaucelle, Stephane Douady, Bruno Moulia, and Herman Hoefte.
\newblock {KymoRod}: a method for automated kinematic analysis of rod-shaped
  plant organs.
\newblock \emph{Plant Journal}, 88\penalty0 (3):\penalty0 468--475, November
  2016.
\newblock ISSN 0960-7412.
\newblock \doi{10.1111/tpj.13255}.
\newblock (doi : 10.1111/tpj.13255).

\bibitem[Bengough et~al.(1997)Bengough, Croser, and
  Pritchard]{bengough_biophysical_1997}
A.~G. Bengough, C.~Croser, and J.~Pritchard.
\newblock A biophysical analysis of root growth under mechanical stress.
\newblock \emph{Plant and Soil}, 189\penalty0 (1):\penalty0 155--164, February
  1997.
\newblock ISSN 0032-079X.
\newblock \doi{10.1023/A:1004240706284}.
\newblock (doi : 10.1023/A:1004240706284).

\bibitem[Bengough et~al.(2011)Bengough, McKenzie, Hallett, and
  Valentine]{bengough_root_2011}
A.~Glyn Bengough, B.~M. McKenzie, P.~D. Hallett, and T.~A. Valentine.
\newblock Root elongation, water stress, and mechanical impedance: a review of
  limiting stresses and beneficial root tip traits.
\newblock \emph{Journal of Experimental Botany}, 62\penalty0 (1):\penalty0
  59--68, January 2011.
\newblock ISSN 0022-0957.
\newblock \doi{10.1093/jxb/erq350}.
\newblock (doi : 10.1093/jxb/erq350).

\bibitem[Bizet et~al.(2016)Bizet, Bengough, Hummel, Bogeat-Triboulot, and
  Dupuy]{bizet_3d_2016}
Francois Bizet, A.~Glyn Bengough, Irene Hummel, Marie-Beatrice
  Bogeat-Triboulot, and Lionel~X. Dupuy.
\newblock {3D} deformation field in growing plant roots reveals both mechanical
  and biological responses to axial mechanical forces.
\newblock \emph{Journal of Experimental Botany}, 67\penalty0 (19):\penalty0
  5605--5614, October 2016.
\newblock ISSN 0022-0957.
\newblock \doi{10.1093/jxb/erw320}.
\newblock (doi : 10.1093/jxb/erw320).

\bibitem[Clark et~al.(1996)Clark, Whalley, Dexter, Barraclough, and
  Leigh]{clark_complete_1996}
L.~J. Clark, W.~R. Whalley, A.~R. Dexter, P.~B. Barraclough, and R.~A. Leigh.
\newblock Complete mechanical impedance increases the turgor of cells in the
  apex of pea roots.
\newblock \emph{Plant Cell and Environment}, 19\penalty0 (9):\penalty0
  1099--1102, September 1996.
\newblock ISSN 0140-7791.
\newblock \doi{10.1111/j.1365-3040.1996.tb00217.x}.
\newblock (doi : 10.1111/j.1365-3040.1996.tb00217.x).

\bibitem[Clark et~al.(1999)Clark, Bengough, Whalley, Dexter, and
  Barraclough]{clark_maximum_1999}
L.~J. Clark, A.~G. Bengough, W.~R. Whalley, A.~R. Dexter, and P.~B.
  Barraclough.
\newblock Maximum axial root growth pressure in pea seedlings: effects of
  measurement techniques and cultivars.
\newblock \emph{Plant and Soil}, 209\penalty0 (1):\penalty0 101--109, 1999.
\newblock ISSN 0032-079X.
\newblock \doi{10.1023/A:1004568714789}.
\newblock (doi : 10.1023/A:1004568714789).

\bibitem[Colombi and Keller(2019)]{colombi_developing_2019}
Tino Colombi and Thomas Keller.
\newblock Developing strategies to recover crop productivity after soil
  compaction-{A} plant eco-physiological perspective.
\newblock \emph{Soil \& Tillage Research}, 191:\penalty0 156--161, August 2019.
\newblock ISSN 0167-1987.
\newblock \doi{10.1016/j.still.2019.04.008}.
\newblock (doi : 10.1016/j.still.2019.04.008).

\bibitem[Cosgrove(1987)]{cosgrove_wall_1987}
Dj~Cosgrove.
\newblock Wall {Relaxation} and the {Driving} {Forces} for {Cell} {Expansive}
  {Growth}.
\newblock \emph{Plant Physiology}, 84\penalty0 (3):\penalty0 561--564, July
  1987.
\newblock ISSN 0032-0889.
\newblock \doi{10.1104/pp.84.3.561}.
\newblock (doi : 10.1104/pp.84.3.561).

\bibitem[Coutand and Moulia(2000)]{coutand_biomechanical_2000}
C.~Coutand and B.~Moulia.
\newblock Biomechanical study of the effect of a controlled bending on tomato
  stem elongation: local strain sensing and spatial integration of the signal.
\newblock \emph{Journal of Experimental Botany}, 51\penalty0 (352):\penalty0
  1825--1842, November 2000.
\newblock ISSN 0022-0957.
\newblock \doi{10.1093/jexbot/51.352.1825}.
\newblock (doi : 10.1093/jexbot/51.352.1825).

\bibitem[Cramer and Bowman(1991)]{cramer_kinetics_1991}
Gr~Cramer and Dc~Bowman.
\newblock Kinetics of {Maize} {Leaf} {Elongation} .1. {Increased} {Yield}
  {Threshold} {Limits} {Short}-{Term}, {Steady}-{State} {Elongation} {Rates}
  {After} {Exposure} to {Salinity}.
\newblock \emph{Journal of Experimental Botany}, 42\penalty0 (244):\penalty0
  1417--1426, November 1991.
\newblock ISSN 0022-0957.
\newblock \doi{10.1093/jxb/42.11.1417}.
\newblock (doi : 10.1093/jxb/42.11.1417).

\bibitem[Dyson et~al.(2014)Dyson, Vizcay-Barrena, Band, Fernandes, French,
  Fozard, Hodgman, Kenobi, Pridmore, Stout, Wells, Wilson, Bennett, and
  Jensen]{dyson_mechanical_2014}
Rosemary~J. Dyson, Gema Vizcay-Barrena, Leah~R. Band, Anwesha~N. Fernandes,
  Andrew~P. French, John~A. Fozard, T.~Charlie Hodgman, Kim Kenobi, Tony~P.
  Pridmore, Michael Stout, Darren~M. Wells, Michael~H. Wilson, Malcolm~J.
  Bennett, and Oliver~E. Jensen.
\newblock Mechanical modelling quantifies the functional importance of outer
  tissue layers during root elongation and bending.
\newblock \emph{New Phytologist}, 202\penalty0 (4):\penalty0 1212--1222, June
  2014.
\newblock ISSN 0028-646X.
\newblock \doi{10.1111/nph.12764}.
\newblock (doi : 10.1111/nph.12764).

\bibitem[Frensch and Hsiao(1994)]{frensch_transient_1994}
J.~Frensch and Tc~Hsiao.
\newblock Transient {Responses} of {Cell} {Turgor} and {Growth} of {Maize}
  {Roots} as {Affected} by {Changes} in {Water} {Potential}.
\newblock \emph{Plant Physiology}, 104\penalty0 (1):\penalty0 247--254, January
  1994.
\newblock ISSN 0032-0889.
\newblock \doi{10.1104/pp.104.1.247}.
\newblock (doi : 10.1104/pp.104.1.247).

\bibitem[Frensch and Hsiao(1995)]{frensch_rapid_1995}
J.~Frensch and Tc~Hsiao.
\newblock Rapid {Response} of the {Yield} {Threshold} and {Turgor} {Regulation}
  {During} {Adjustment} of {Root}-{Growth} to {Water}-{Stress} in {Zea}-{Mays}.
\newblock \emph{Plant Physiology}, 108\penalty0 (1):\penalty0 303--312, May
  1995.
\newblock ISSN 0032-0889.
\newblock \doi{10.1104/pp.108.1.303}.
\newblock (doi : 10.1104/pp.108.1.303).

\bibitem[Gardiner et~al.(2016)Gardiner, Berry, and
  Moulia]{gardiner_review_2016}
Barry Gardiner, Peter Berry, and Bruno Moulia.
\newblock Review: {Wind} impacts on plant growth, mechanics and damage.
\newblock \emph{Plant Science}, 245:\penalty0 94--118, April 2016.
\newblock ISSN 0168-9452.
\newblock \doi{10.1016/j.plantsci.2016.01.006}.
\newblock (doi : 10.1016/j.plantsci.2016.01.006).

\bibitem[Gill and Bolt(1955)]{gill_pfeffers_1955}
W.~Gill and G.~Bolt.
\newblock Pfeffer's {Studies} of the {Root} {Growth} {Pressures} {Exerted} by
  {Plants}.
\newblock 1955.
\newblock \doi{10.2134/AGRONJ1955.00021962004700040004X}.
\newblock (doi : 10.2134/AGRONJ1955.00021962004700040004X).

\bibitem[Greacen and Oh(1972)]{greacen_physics_1972}
El~Greacen and Js~Oh.
\newblock Physics of {Root} {Growth}.
\newblock \emph{Nature-New Biology}, 235\penalty0 (53):\penalty0 24--\&, 1972.
\newblock \doi{10.1038/newbio235024a0}.
\newblock (doi : 10.1038/newbio235024a0).

\bibitem[Green et~al.(1971)Green, Erickson, and Buggy]{green_metabolic_1971}
Pb~Green, Ro~Erickson, and J.~Buggy.
\newblock Metabolic and {Physical} {Control} of {Cell} {Elongation} {Rate} -
  in-{Vivo} {Studies} in {Nitella}.
\newblock \emph{Plant Physiology}, 47\penalty0 (3):\penalty0 423--\&, 1971.
\newblock ISSN 0032-0889.
\newblock \doi{10.1104/pp.47.3.423}.
\newblock (doi : 10.1104/pp.47.3.423).

\bibitem[Gregory(2006)]{gregory_plant_2006}
Peter~J. Gregory.
\newblock \emph{Plant {Roots}: {Growth}, {Activity} and {Interactions} with the
  {Soil}}.
\newblock Blackwell Publishing Ltd, 2006.
\newblock ISBN 978-1-4051-1906-1.
\newblock (doi : 10.1002/9780470995563).

\bibitem[Gregory et~al.(2009)Gregory, Bengough, Grinev, Schmidt, Thomas,
  Wojciechowski, and Young]{gregory_root_2009}
Peter~J. Gregory, A.~Glyn Bengough, Dmitri Grinev, Sonja Schmidt, W.~Thomas,
  Tobias Wojciechowski, and Iain~M. Young.
\newblock Root phenomics of crops: opportunities and challenges.
\newblock \emph{Functional Plant Biology}, 36\penalty0 (10-11):\penalty0
  922--929, 2009.
\newblock ISSN 1445-4408.
\newblock \doi{10.1071/FP09150}.
\newblock (doi : 10.1071/FP09150).

\bibitem[Griffiths et~al.(2022)Griffiths, Delory, Jawahir, Wong, Bagnall, Dowd,
  Nusinow, Miller, and Topp]{griffiths_optimisation_2022}
Marcus Griffiths, Benjamin~M. Delory, Vanessica Jawahir, Kong~M. Wong, G.~Cody
  Bagnall, Tyler~G. Dowd, Dmitri~A. Nusinow, Allison~J. Miller, and
  Christopher~N. Topp.
\newblock Optimisation of root traits to provide enhanced ecosystem services in
  agricultural systems: {A} focus on cover crops.
\newblock \emph{Plant Cell and Environment}, 45\penalty0 (3):\penalty0
  751--770, March 2022.
\newblock ISSN 0140-7791.
\newblock \doi{10.1111/pce.14247}.
\newblock (doi : 10.1111/pce.14247).

\bibitem[Jin et~al.(2013)Jin, Shen, Ashton, Dodd, Parry, and
  Whalley]{jin_how_2013}
Kemo Jin, Jianbo Shen, Rhys~W. Ashton, Ian~C. Dodd, Martin A.~J. Parry, and
  William~R. Whalley.
\newblock How do roots elongate in a structured soil?
\newblock \emph{Journal of Experimental Botany}, 64\penalty0 (15):\penalty0
  4761--4777, November 2013.
\newblock ISSN 0022-0957.
\newblock \doi{10.1093/jxb/ert286}.

\bibitem[Kolb et~al.(2017)Kolb, Legue, and
  Bogeat-Triboulot]{kolb_physical_2017}
Evelyne Kolb, Valerie Legue, and Marie-Beatrice Bogeat-Triboulot.
\newblock Physical root-soil interactions.
\newblock \emph{Physical Biology}, 14\penalty0 (6):\penalty0 065004, December
  2017.
\newblock ISSN 1478-3967.
\newblock \doi{10.1088/1478-3975/aa90dd}.
\newblock (doi : 10.1088/1478-3975/aa90dd).

\bibitem[Landrein and Ingram(2019)]{landrein_connected_2019}
Benoit Landrein and Gwyneth Ingram.
\newblock Connected through the force: mechanical signals in plant development.
\newblock \emph{Journal of Experimental Botany}, 70\penalty0 (14):\penalty0
  3507--3519, July 2019.
\newblock ISSN 0022-0957.
\newblock \doi{10.1093/jxb/erz103}.
\newblock (doi : 10.1093/jxb/erz103).

\bibitem[Lockhart(1965)]{lockhart_analysis_1965}
Ja~Lockhart.
\newblock An {Analysis} of {Irreversible} {Plant} {Cell} {Elongation}.
\newblock \emph{Journal of Theoretical Biology}, 8\penalty0 (2):\penalty0
  264--\&, 1965.
\newblock ISSN 0022-5193.
\newblock \doi{10.1016/0022-5193(65)90077-9}.
\newblock (doi : 10.1016/0022-5193(65)90077-9).

\bibitem[Lynch et~al.(2022)Lynch, Mooney, Strock, and
  Schneider]{lynch_future_2022}
Jonathan~P. Lynch, Sacha~J. Mooney, Christopher~F. Strock, and Hannah~M.
  Schneider.
\newblock Future roots for future soils.
\newblock \emph{Plant Cell and Environment}, 45\penalty0 (3):\penalty0
  620--636, March 2022.
\newblock ISSN 0140-7791.
\newblock \doi{10.1111/pce.14213}.
\newblock (doi : 10.1111/pce.14213).

\bibitem[Nonami and Boyer(1990{\natexlab{a}})]{nonami_primary_1990}
H.~Nonami and Js~Boyer.
\newblock Primary {Events} {Regulating} {Stem} {Growth} at {Low} {Water}
  {Potentials}.
\newblock \emph{Plant Physiology}, 93\penalty0 (4):\penalty0 1601--1609, August
  1990{\natexlab{a}}.
\newblock ISSN 0032-0889.
\newblock \doi{10.1104/pp.93.4.1601}.
\newblock (doi : 10.1104/pp.93.4.1601).

\bibitem[Nonami and Boyer(1990{\natexlab{b}})]{nonami_wall_1990}
H.~Nonami and Js~Boyer.
\newblock Wall {Extensibility} and {Cell} {Hydraulic} {Conductivity} {Decrease}
  in {Enlarging} {Stem} {Tissues} at {Low} {Water} {Potentials}.
\newblock \emph{Plant Physiology}, 93\penalty0 (4):\penalty0 1610--1619, August
  1990{\natexlab{b}}.
\newblock ISSN 0032-0889.
\newblock \doi{10.1104/pp.93.4.1610}.
\newblock (doi : 10.1104/pp.93.4.1610).

\bibitem[Peters and Baskin(2006)]{peters_tailor-made_2006}
Winfried~S. Peters and Tobias~I. Baskin.
\newblock Tailor-made composite functions as tools in model choice: the case of
  sigmoidal vs bi-linear growth profiles.
\newblock \emph{Plant Methods}, 2:\penalty0 11, 2006.
\newblock \doi{10.1186/1746-4811-2-11}.
\newblock (doi : 10.1186/1746-4811-2-11).

\bibitem[Pfeffer(1893)]{pfeffer_druck-_1893}
W.~Pfeffer.
\newblock \emph{Druck- und {Arbeitsleistung} durch wachsende {Pflanzen}}.
\newblock S. Hirzel,, Leipzig,, 1893.
\newblock ISBN DOI: 10.5962/bhl.title.13148.
\newblock (doi : 10.5962/bhl.title.13148).

\bibitem[Popova et~al.(2016)Popova, van Dusschoten, Nagel, Fiorani, and
  Mazzolai]{popova_plant_2016}
Liyana Popova, Dagmar van Dusschoten, Kerstin~A. Nagel, Fabio Fiorani, and
  Barbara Mazzolai.
\newblock Plant root tortuosity: an indicator of root path formation in soil
  with different composition and density.
\newblock \emph{Annals of Botany}, 118\penalty0 (4):\penalty0 685--698, October
  2016.
\newblock ISSN 0305-7364.
\newblock \doi{10.1093/aob/mcw057}.
\newblock (doi : 10.1093/aob/mcw057).

\bibitem[Potocka and Szymanowska-Pulka(2018)]{potocka_morphological_2018}
Izabela Potocka and Joanna Szymanowska-Pulka.
\newblock Morphological responses of plant roots to mechanical stress.
\newblock \emph{Annals of Botany}, 122\penalty0 (5):\penalty0 711--723, October
  2018.
\newblock ISSN 0305-7364.
\newblock \doi{10.1093/aob/mcy010}.
\newblock (doi : 10.1093/aob/mcy010).

\bibitem[Pritchard(1994)]{pritchard_control_1994}
J.~Pritchard.
\newblock The {Control} of {Cell} {Expansion} in {Roots}.
\newblock \emph{New Phytologist}, 127\penalty0 (1):\penalty0 3--26, May 1994.
\newblock ISSN 0028-646X.
\newblock \doi{10.1111/j.1469-8137.1994.tb04255.x}.
\newblock (doi : 10.1111/j.1469-8137.1994.tb04255.x).

\bibitem[Rellan-Alvarez et~al.(2016)Rellan-Alvarez, Lobet, and
  Dinneny]{rellan-alvarez_environmental_2016}
Ruben Rellan-Alvarez, Guillaume Lobet, and Jose~R. Dinneny.
\newblock Environmental {Control} of {Root} {System} {Biology}.
\newblock In S.~S. Merchant, editor, \emph{Annual {Review} of {Plant}
  {Biology}, {Vol} 67}, volume~67, pages 619--642. Annual Reviews, Palo Alto,
  2016.
\newblock ISBN 978-0-8243-0667-0.
\newblock (doi : 10.1146/annurev-arplant-043015-111848).

\bibitem[Sharp et~al.(1988)Sharp, Silk, and Hsiao]{sharp_growth_1988}
Re~Sharp, Wk~Silk, and Tc~Hsiao.
\newblock Growth of the {Maize} {Primary} {Root} at {Low} {Water} {Potentials}
  .1. {Spatial}-{Distribution} of {Expansive} {Growth}.
\newblock \emph{Plant Physiology}, 87\penalty0 (1):\penalty0 50--57, May 1988.
\newblock ISSN 0032-0889.
\newblock \doi{10.1104/pp.87.1.50}.
\newblock (doi : 10.1104/pp.87.1.50).

\bibitem[Silk and Erickson(1979)]{silk_kinematics_1979}
Wk~Silk and Ro~Erickson.
\newblock Kinematics of {Plant}-{Growth}.
\newblock \emph{Journal of Theoretical Biology}, 76\penalty0 (4):\penalty0
  481--501, 1979.
\newblock ISSN 0022-5193.
\newblock \doi{10.1016/0022-5193(79)90014-6}.
\newblock (doi : 10.1016/0022-5193(79)90014-6).

\bibitem[Souty(1987)]{souty_mechanical-behavior_1987}
N.~Souty.
\newblock Mechanical-{Behavior} of {Growing} {Roots} .1. {Measurement} of
  {Penetration} {Force}.
\newblock \emph{Agronomie}, 7\penalty0 (8):\penalty0 623--630, 1987.
\newblock ISSN 0249-5627.
\newblock \doi{10.1051/agro:19870810}.
\newblock (doi : 10.1051/agro:19870810).

\bibitem[Stubbs et~al.(2019)Stubbs, Cook, and Niklas]{stubbs_general_2019}
Christopher~J. Stubbs, Douglas~D. Cook, and Karl~J. Niklas.
\newblock A general review of the biomechanics of root anchorage.
\newblock \emph{Journal of Experimental Botany}, 70\penalty0 (14):\penalty0
  3439--3451, July 2019.
\newblock ISSN 0022-0957.
\newblock \doi{10.1093/jxb/ery451}.
\newblock (doi : 10.1093/jxb/ery451).

\bibitem[Tracy et~al.(2011)Tracy, Black, Roberts, and Mooney]{tracy_soil_2011}
Saoirse~R. Tracy, Colin~R. Black, Jeremy~A. Roberts, and Sacha~J. Mooney.
\newblock Soil compaction: a review of past and present techniques for
  investigating effects on root growth.
\newblock \emph{Journal of the Science of Food and Agriculture}, 91\penalty0
  (9):\penalty0 1528--1537, July 2011.
\newblock ISSN 0022-5142.
\newblock \doi{10.1002/jsfa.4424}.
\newblock (doi : 10.1002/jsfa.4424).

\bibitem[Veen and Boone(1990)]{veen_influence_1990}
Bw~Veen and Fr~Boone.
\newblock The {Influence} of {Mechanical} {Resistance} and {Soil}-{Water} on
  the {Growth} of {Seminal} {Roots} of {Maize}.
\newblock \emph{Soil \& Tillage Research}, 16\penalty0 (1-2):\penalty0
  219--226, April 1990.
\newblock ISSN 0167-1987.
\newblock \doi{10.1016/0167-1987(90)90031-8}.
\newblock (doi : 10.1016/0167-1987(90)90031-8).

\bibitem[Walter et~al.(2003)Walter, Feil, and Schurr]{walter_expansion_2003}
A.~Walter, R.~Feil, and U.~Schurr.
\newblock Expansion dynamics, metabolite composition and substance transfer of
  the primary root growth zone of {Zea} mays {L}. grown in different external
  nutrient availabilities.
\newblock \emph{Plant Cell and Environment}, 26\penalty0 (9):\penalty0
  1451--1466, September 2003.
\newblock ISSN 0140-7791.
\newblock \doi{10.1046/j.0016-8025.2003.01068.x}.
\newblock (doi : 10.1046/j.0016-8025.2003.01068.x).

\bibitem[Youssef et~al.(2018)Youssef, Bizet, Bastien, Legland,
  Bogeat-Triboulot, and Hummel]{youssef_quantitative_2018}
Chvan Youssef, Francois Bizet, Renaud Bastien, David Legland, Marie-Beatrice
  Bogeat-Triboulot, and Irene Hummel.
\newblock Quantitative dissection of variations in root growth rate: a matter
  of cell proliferation or of cell expansion?
\newblock \emph{Journal of Experimental Botany}, 69\penalty0 (21):\penalty0
  5157--5168, October 2018.
\newblock ISSN 0022-0957.
\newblock \doi{10.1093/jxb/ery272}.
\newblock (doi : 10.1093/jxb/ery272).

\bibitem[Zhu and Boyer(1992)]{zhu_enlargement_1992}
Gl~Zhu and Js~Boyer.
\newblock Enlargement in {Chara} {Studied} with a {Turgor} {Clamp} -
  {Growth}-{Rate} {Is} {Not} {Determined} by {Turgor}.
\newblock \emph{Plant Physiology}, 100\penalty0 (4):\penalty0 2071--2080,
  December 1992.
\newblock ISSN 0032-0889.
\newblock \doi{10.1104/pp.100.4.2071}.
\newblock (doi : 10.1104/pp.100.4.2071).

\end{thebibliography}
\end{document}